\documentclass[runningheads]{svmult}

\usepackage{psfig}

\usepackage{makeidx}   
\usepackage{graphicx}  
\usepackage{subeqnar}  
\usepackage{multicol}  
\usepackage{physprbb}  
\makeindex             

\begin{document}
\title*{SIMULATION OF MODELS FOR THE \protect\newline GLASS TRANSITION:
IS THERE PROGRESS?}
\toctitle{SIMULATION OF MODELS FOR THE GLASS TRANSITION: \protect\newline IS THERE PROGRESS}
%
%
\titlerunning{SIMULATION OF MODELS FOR THE GLASS TRANSITION}
%
\author{Kurt Binder\inst{1}
\and J\"org Baschnagel\inst{2} \and Walter Kob\inst{3} \and Wolfgang Paul\inst{1} }
\authorrunning{Kurt Binder et al.}
%
%
\institute{Institut f\"ur Physik, Johannes-Gutenberg-Universit\"at Mainz, 
D-55099 Mainz, Germany 
\and Institut Charles Sadron, Universit\'e Louis Pasteur, 6, rue Bossingault, 
F-67083 Strasbourg, France \and 
Laboratoire des Verres, Universit\'e Montpellier II, F-34095 Montpellier, France}

\maketitle              

\begin{abstract}
The glass transition of supercooled fluids is a particular challenge for
computer simulation, because the (longest) relaxation times increase by
about 15 decades upon approaching the transition temperature $T_{g}$.
Brute-force molecular dynamics simulations, as presented here for molten
SiO${_2}$ and coarse-grained bead-spring models of polymer chains, can
yield very useful insight about the first few decades of this slowing
down. Hence this allows to access the temperature range around $T_c$
of the so-called mode coupling theory, whereas the dynamics around the
experimental glass transition is completely out of reach.  While methods
such as ``parallel tempering'' improve the situation somewhat, a method
that allows to span a significant part of the region $T_{g}\leq T\leq
T_{c}$ is still lacking.  Only for abstract models such as the infinite
range $10$-state Potts glass with a few hundred spins this region can
be explored.  However this model suffers from very strong finite size
effects thus making it difficult to extrapolate the results obtained
for the finite system sizes to the thermodynamic limit.

For the case of polymer melts, two different strategies to use lattice
models instead of continuum models are discussed: In the first approach,
a mapping of an atomistically realistic model of polyethylene to
the bond fluctuation model with suitable effective potentials and
a temperature-dependent time rescaling factor is attempted. In the
second approach, devoted to a test of the entropy theory, moves that are
artificial but which lead to a faster relaxation (``slithering snake''
algorithm) are used, to get at least static properties at somewhat lower
temperatures than possible with a ``realistic'' dynamics. The merits
and shortcomings of all these approaches are discussed.

\end{abstract}

\section{Introduction}
The reason for the slowing down of the dynamics of supercooled liquids
and the resulting glass transition to an amorphous solid is one of
the biggest unsolved problems in the physics of condensed matter
\cite{1,2,3,4,5} and it is also a particular challenge for computer
simulation \cite{6,7,8,9,10,11,62}. The present introductory section
intends to remind the reader on the main experimental facts and some
theoretical ideas about the glass transition, and will also serve
to make clear why in this problem there exists a  gap of time-scales
that simulations need to bridge.

\begin{figure}
\centerline{
\psfig{figure=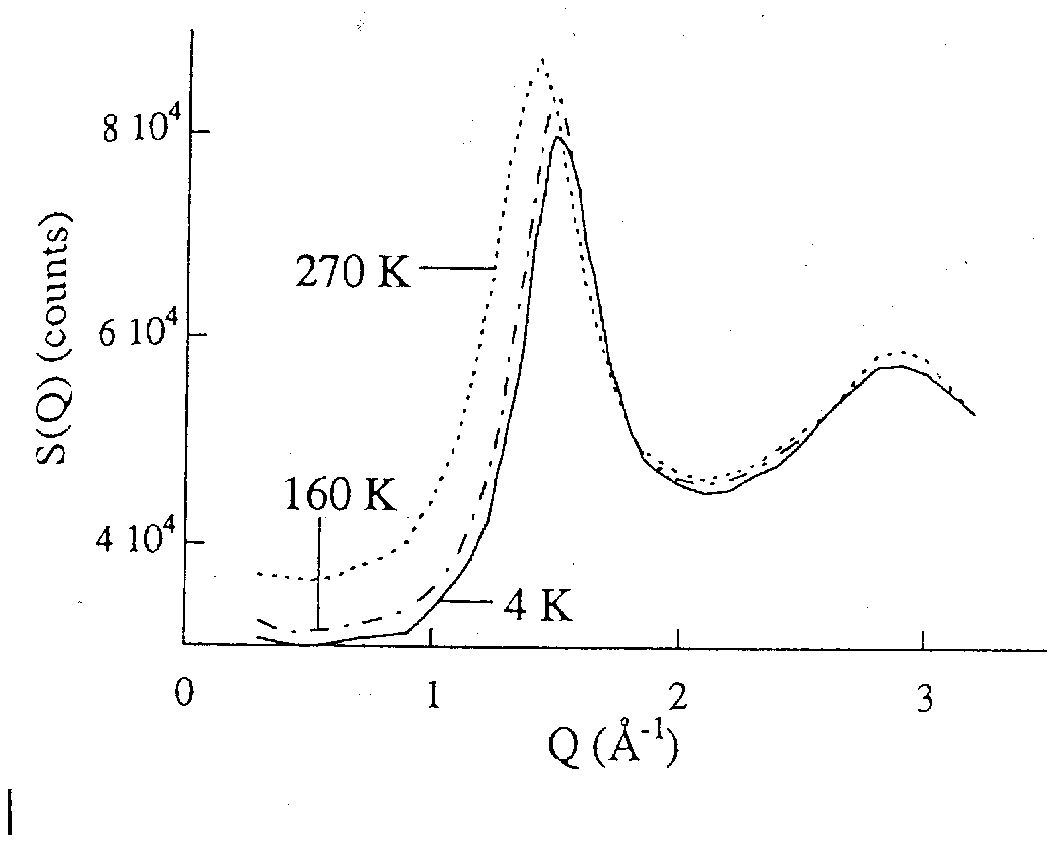,width=11.0cm,height=7.0cm}}
\centerline{
\psfig{figure=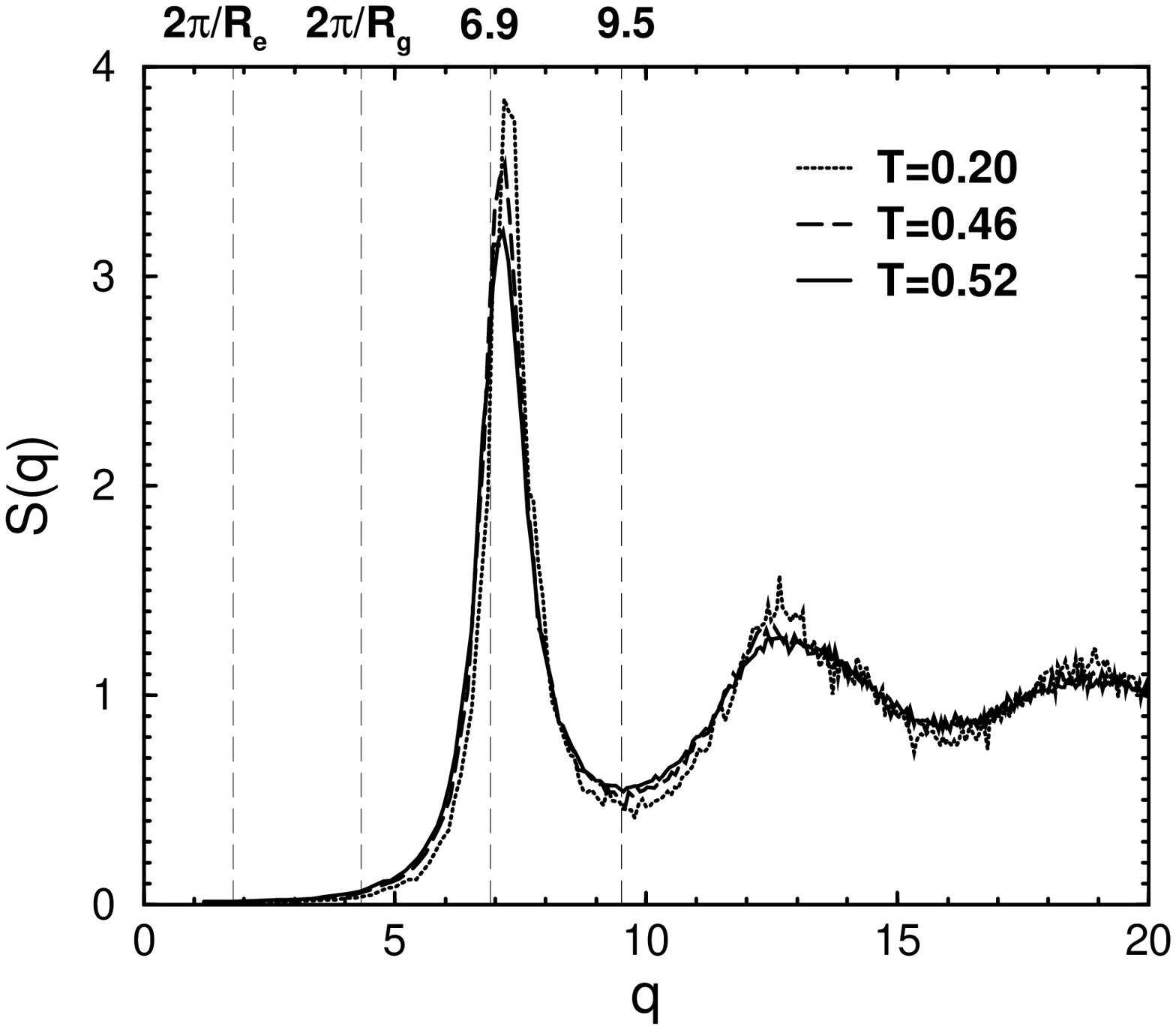,width=10.0cm,height=7.0cm}
}
\caption{\label{fig1}a) Static collective structure factor of
polybutadiene at temperatures $T=4$K, $T=16$K, and $T=270$K. Note that
for this system the glass transition temperature is $T_g=180K$ and the
critical temperature of mode coupling theory \cite{3} is $T_c=220$K.
The scattering background is not subtracted here, thus the zero of
the ordinate axis is not known precisely, and the ordinate units are
just measuring absolute scattering intensities. From Arbe {\it et al.}
\cite{12}. b) Static collective structure factor $S(q)$ plotted versus
wave-vector $q$, for a bead-spring model of flexible polymer chains
with chain length $N=10$. Beads interact with the potential given in
Eqs.~(\ref{eq9})-(\ref{eq11}).  and lengths are measured in units of
$\sigma$, temperatures in units of $\varepsilon$. Three temperatures
$T=0.2$, $0.46$ and $0.52$ are shown (note that $T_g \approx 0.41$
and $T_c \approx 0.45$ for this model). The vertical lines highlight
characteristic inverse length scales (related to the end-to-end distance
$R_e$ and radius of gyration $R_g$ of the chains as
well as the first maximum and minimum of $S(q)$). From Baschnagel el
at. \cite{13}.} 
\end{figure}

\begin{figure}
\centerline{
\psfig{figure=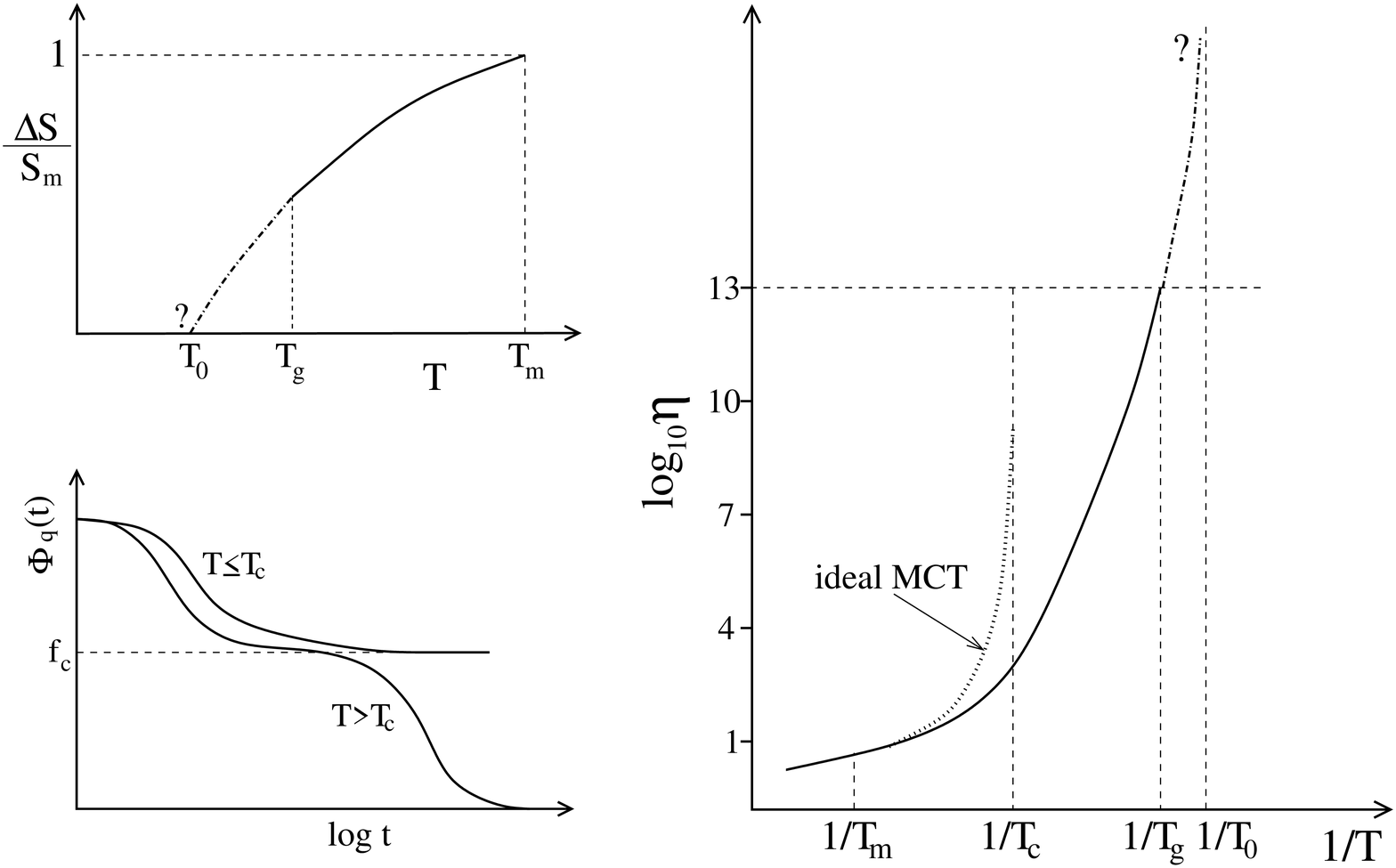,width=12.0cm,height=7.0cm}
}
\caption{\label{fig2}Right figure: Schematic plot of the viscosity
$\eta(T)$ of a fluid (note $\eta (T) \propto \tau$) vs. inverse
temperature $1/T$. The location of the melting temperature $(T_m)$,
the critical temperature of mode coupling theory $(T_c)$ \cite{3}, the
glass transition temperature $(T_g)$ and the Vogel-Fulcher-Kauzmann
temperature \cite{1,14} $(T_0)$ are shown on the abscissa. The glass
transition temperature $T_g$ is defined empirically requiring \cite{1}
$ \eta(T=T_g)=10^{13}$ Poise. Two complementary concepts to explain
the glass transition are indicated by the schematic plots on the left:
The lower figure shows the time correlation function $\Phi_q(t)$ for
density fluctuations at wave-vector $q$ which according to idealized mode
coupling theory shows at a temperature $T_c$ a nonzero ``non-ergodicity
parameter'' $f_c$ \cite{3}. For $T$ somewhat larger than $T_c$, $\Phi_q
(t)$ exhibits a plateau and the ``lifetime'' $\tau$ of this plateau
(as well as $\eta$) diverge as one approaches $T_c$ \cite{3}. The upper
figure shows the entropy difference $\Delta S(T)=S_{\rm fluid}-S_{\rm
crystal}$, with $S_m \equiv \Delta S(T_m)$. The linear extrapolation
of $\Delta S$ for $T<T_g$ defines the Kauzmann temperature $T_0$ via $
\Delta S(T=T_0)=0$ \cite{14}. Adapted from Binder \cite{15}.}

\end{figure}

As is well known, it is already a problem to characterize the static
structure of a glass: the structure of an amorphous material is not
regular like a crystalline solid, but shows only short range order
similar to a liquid. However, the latter flows, while the amorphous
solid is rigid! In fact, if one makes a scattering experiment,
it is hard to distinguish from the structure whether one has a fluid
above the glass transition temperature $T_g$ or a solid below $T_g$
(Fig.~\ref{fig1}) \cite{12,13}. If one approaches $T_g$, the structural
relaxation time $\tau$ which is related to the viscosity $\eta (T)$, for
instance - increases smoothly by up to 15 decades, as shown schematically
in Fig.~\ref{fig2}, without any accompanying significant structural change
detectable by scattering experiments (Fig.~\ref{fig1}). This increase
of $\eta(T)$ is often fitted to the Vogel-Fulcher relation \cite{1}

\begin{equation} 
\eta(T) \propto \exp[E_{VF}/(T-T_{VF})] \; ,
\label{eq1}
\end{equation}

\noindent
where $E_{VF}$ is an effective activation barrier. From this functional
form it is clear that $\eta (T)$ is predicted to diverge at the
Vogel-Fulcher temperature $T_{VF}$, which is lower than $T_g$, of course,
if one invokes the empirical definition of $T_g$ via $\eta(T=T_g)=10^{13}$
Poise \cite{1}. However, it is questionable whether the temperature
dependence of $\eta$ as given by Eq.~(\ref{eq1}) really holds. 

Another very common relation often used to describe various relaxation
functions of glassforming fluids is the Kohlrausch-Williams-Watts
function, also called stretched exponential, \cite{1},

\begin{equation} 
\varphi(t) \propto \exp[-(t/\tau)^\beta ] \; .
\label{eq2}
\end{equation}

\noindent
This relation involves an exponent $\beta \leq 1$, whose precise physical
significance is somewhat obscure. Again it is unknown under which
circumstances (if any) Eq.~(\ref{eq2}) is exact, and in which it is just
a convenient fitting formula to represent data.

Often it is claimed that the glass transition is a purely kinetic
phenomenon, and if one would be able to wait long enough (which could
mean times like the age of the universe, however!) one could see that
glass is not really a solid but still a fluid that flows. However, this
idea is not generally accepted, since there are some indications that
there may be an underlying quasi-equilibrium phase transition between
metastable phases, namely from the supercooled metastable fluid to an
(also metastable) ideal glass phase (the stable phase for temperatures
lower than the melting temperature $T_m$ is of course the crystal). Such
an indication is Kauzmann's entropy paradox \cite{14}: By studying the
difference in entropy between liquid and crystal one finds that near
$T_g$ the difference $\Delta S(T)=S_{\rm fluid}-S_{\rm crystal}$ has
decreased to about 1/3 of its value $S_m$ at the melting/crystallization
temperature $T_m$. If this trend is extrapolated (linearly in $T$)
to even lower temperatures, $\Delta S(T)$ would become negative
below the Kauzmann temperature $T_0$ (which is usually quite close
to the Vogel-Fulcher-temperature $T_{VF}$), see Fig.~\ref{fig2}. It
would indeed be paradox if the entropy of the supercooled fluid (with
its disordered structure) were less than the entropy of the ordered
solid! One possibility to bypass the problem is to assume that this
``entropy catastrophe'' is avoided by a phase transition at $T_0$ (or
at some temperature in between $T_0$ and $T_g$). In fact, for the glass
transition of polymer melts Gibbs and Di Marzio \cite{16} proposed an
approximate theory that shows such a vanishing of the entropy at $T_0$,
and subsequently Adam and Gibbs \cite{17} suggested arguments to show
that Eq.~(\ref{eq1}) holds with $T_{VF}=T_0$. However, although these
concepts enjoy some popularity, all these arguments are based on very
crude and hardly justifiable assumptions and approximations, and hence
they are not accepted by many researchers. For instance, the mode coupling
theory of the glass transition (MCT)\cite{3} claims that there is indeed
an underlying transition but this is {\it not} a phase transition in
the sense of thermodynamics but rather a ``dynamical transition'' from
an ergodic to a nonergodic behavior. This transition should occur at
a critical temperature $T_c$ and can be seen in the form of the time
dependence of the correlation function of density fluctuations or
its Fourier transform $\Phi_q(T)$, see Fig.~\ref{fig2}. Above $T_c$,
this correlator decays to zero, but as $T_c$ is approached a plateau
develops whose ``lifetime'' gets larger and larger until it diverges,
in the ideal case: the system gets ``stuck'', the decay of $\Phi_q(t)$
stops at the ``nonergodicity parameter'' $f_c$, an order parameter for
the glass transition that appears discontinuously at $T_c$. The physical
idea behind this theory is the ``cage picture'': the motion of any atom
in a dense fluid is constrained by its neighbors, which form a cage
around it. At low enough temperatures the escape out of the cage gets
blocked. MCT predicts that close to this dynamical transition $\tau$
and $\eta(T)$ show a power-law divergence as one approaches $T_c$,

\begin{equation}
\tau \propto \eta (T) \propto  (T-T_c)^{-\gamma} \, .
\label{eq3}
\end{equation}

\noindent
In reality this dependence is, however, observed only in a limited
temperature interval. The way out of this dilemma is the argument
that one must not neglect (as ``idealized'' mode coupling theory does
\cite{3}) thermally activated processes, so-called ``hopping processes'',
by which atoms supposedly can escape from their cage when $T$ is less
than $T_c$. The theory then claims \cite{18} that a simple Arrhenius
behavior results in this region, $\log \tau \propto 1/T$ for $T<T_c$,
and in the vicinity of $T_c$ the power-law divergence of Eq.~(\ref{eq3})
is rounded off to a smooth crossover from the power law to the Arrhenius
divergence. Thus this theory does not involve any phase transition,
there is just a smooth crossover from one type of dynamical behavior
to another one near $T_c$, and $T_g$ means that relaxation times have
grown so large that the system falls out of equilibrium.

In real systems this crossover seems to occur at a viscosity somewhere
between $10$ and $10^{3}$ Poise, i.e. a time window that can be explored
with molecular dynamics (MD) simulations. Hence such simulations are able
to investigate the beginning of the approach to the critical temperature
$T_c$ which MCT describes \cite{3} and hence are very useful to check
the validity of this theory. However, the following $10$ decades of the
viscosity between $T_c$ and $T_g$ are out of reach for MD simulations
so far. Unfortunately this is precisely the region that one needs to
explore, for a definite distinction between the theories!

Thus although straightforward atomistic MD methods \cite{19,20} clearly
face a dilemma, we shall nevertheless describe how far one can push this
type of approach, choosing SiO$_2$ as an example (Sec. 2). A method
to extend the range of times and accessible temperatures somewhat is
the concept of ``parallel tempering'' \cite{21a,21,21b,21c,22}, and this approach
and its problems will be presented in Sec. 3. For the sake of contrast,
Sec. 4 will then describe the $10$-state Potts glass model. Although this
model is only an abstract caricature for a real glass, it has the merit
that quite a few results are known analytically and that Monte Carlo
simulations are possible at $T_c$ and even at lower temperatures, if one
considers only systems of a few hundred Potts spins. The disappointing
aspect is, Sec. 4, that even in this very idealized case one learns
relatively little about the glass transition of the infinite system, since
one has to fight against dramatic finite size effects \cite{23,24,25,26}!

Then we shall describe briefly (Sec. 5) a coarse-grained model of
short polymer chains \cite{13,27,28,29,30,31,32,33}. This beadspring
model is quite successful in reproducing a number of experimental
results qualitatively, as already exemplified in Fig.~\ref{fig1}. The
cooling rates that one can reach are about 3 orders of magnitude
smaller than for SiO$_2$. Thus the model is very useful for testing
mode coupling theory \cite{32,33}. However, also for this system
there is actually only a dim hope that one can get distinctly below
the critical temperature $T_c$! With respect to that it may be
better to work with the so-called bond fluctuation model on the
lattice \cite{6,34,35,36,37,38,39,40,41,42,43} - a system which
allows to equilibrate melts at low temperatures with artificial moves
\cite{40,41,42,43}. By simulations of this model also the configurational
entropy and its temperature dependence can be extracted \cite{40,41}
and thus it can be shown that the entropy theory of Gibbs and Di
Marzio \cite{16} is rather inaccurate and misleading (Sec. 6). Finally,
an interesting variant (Sec. 7) of the bond fluctuation model will
be considered. Here one uses effective potentials that are constructed
such that a real material is mimicked, e.g. polyethylene \cite{44}. This
trick allows that part of the problem of bridging the time scales is taken
care of by a ``time rescaling factor'' \cite{44}, a special translation
factor between the physical time and the Monte Carlo time.  Sec. 8 then
will summarize some of the conclusions emerging from all this work.

\section{Towards the simulation of real glassy materials: The case of SiO${_2}$}

Molten SiO${_2}$ is a prototype of a network glassformer. Furthermore it
is a system that is well suited for molecular dynamics simulations since
a very well-tested pair potential based on quantum-chemical calculation
has been developed \cite{45}. By a suitable combination of long-range
Coulomb interactions and short range forces, chosen in the form

\begin{equation}
V_{ij}(r)= \frac{q_i q_j e^2}{r} + A_{ij} \exp (-B_{ij} r) - 
C_{ij} / r^6 \qquad \mbox{with } i,j \in  \{\mbox{Si,O}\},
\label{eq4}
\end{equation}

\noindent
the effective interaction between the ions can be described reliably.
Here $e$ is the charge of an electron, $q_{\rm O}=-1.2$, $q_{\rm Si}=2.4$,
and the values of the parameters $A_{ij}$, $B_{ij}$ and $C_{ij}$ can
be found in Ref.~\cite{45}.  This potential is able to describe the
formation of covalent bonds without the explicit assumption of three-body
forces, whose calculation would be very time consuming. Due to the long range of
the electrostatic interactions, Ewald summation techniques have to be
used, while the short range part of the potential can be cut off at a
suitable radius $r_c$. It turns out that $r_c=5.5$\AA\,
yields good results~\cite{46}. The MD time step, however, must be chosen
relatively small, namely $\delta t=1.6$fs. Note that the presence of
the long range Coulombic interactions make the calculation of the forces
still a quite CPU-intensive task.  Furthermore one has to average the
results over several independent runs in order to improve the statistics.

In a first set of simulations, the method used to cool the sample was very
similar to the procedure used in glass factories where the temperature is
reduced linearly with time $t$, starting from some initial temperature
$T_i$ such that $T(t)=T_i- \gamma t$, where $\gamma$ is the cooling
rate \cite{46}. The main difference between the simulation and the
cooling of the real material are the actual numbers used here: The
initial temperature that had to be chosen in the simulation was very
high, $T_i=7000$K, and cooling rates were between $\gamma=10^{15}$K/s
and $\gamma=10^{12}$K/s \cite{46}. In contrast the glass factory uses
typical initial temperatures of 1600K and cooling rates of $1$K/s or even
less, so the simulation is at least 12 orders of magnitude off! Despite
these extremely high cooling rates - which are inevitable due to the
heavy computational burden - the generated structures are qualitatively
reasonable.  In particular one obtains random tetrahedral networks in
which almost all Si atom sit in the center of a tetrahedron and most of
the O atoms sit at the corners.

\begin{figure}
\centerline{
\psfig{figure=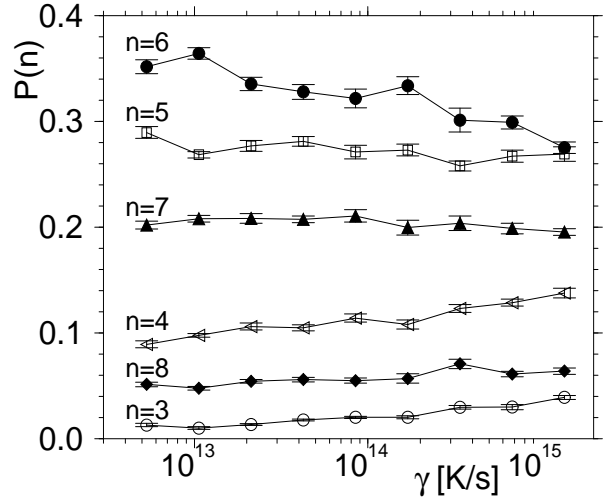,width=8.0cm,height=6.5cm}
}
\caption{\label{fig3}Cooling rate dependence of the probability $P(n)$
that in the network structure of SiO$_2$ a ring of size $n$ is present. A
ring is defined as the shortest connection of consecutive Si--O elements
that form a closed loop and $n$ is the number of these segments. In
this simulation we used $668$ oxygen and $334$ Si-atoms, and cooled the
sample at constant pressure $p=0$ in an $NpT$ simulation, cooling from
the initial temperature $T_i=7000$K to the final temperature $T=0$ K.
An average over $10$ independent runs was performed, allowing to
estimate the statistical errors given in the figure. From Vollmayr {\it et
al.} \cite{46}.}
\end{figure}

Earlier investigators (for a review see \cite{7,8,46}) were so bold
to claim that such glass structures are identical to those occurring in
nature, denying any significant dependence on cooling rate. However, as we
have shown \cite{46}, such a claim is foolish since one sees a pronounced
dependence on cooling rate in many quantities, including the structure. As
a typical example we show in  Fig.~\ref{fig3} how the distribution of
the length $n$ of rings depends on $\gamma$\cite{46}. (See the figure
caption for a definition of this length.) It is seen that over the range
of $\gamma$ that is accessible there is a significant increase of $P(n=6)$
and a significant decrease of $P(n=3)$ and $P(n=4)$, while $P(n=5)$,
$P(n=7)$ and $P(n=8)$ almost stay constant. Clearly an extrapolation of
such data to the physically relevant cooling rate $\gamma=1$K/s is very
difficult, and perhaps not yet possible: Perhaps $P(n=3)$ and $P(n=4)$
are already practically equal to zero for $\gamma=1$K/s - we don't
really know. Even simple quantities, such as the density of the glass
at low temperatures, are hard to predict reliably. (This problem is
also complicated by the fact that molten SiO$_2$ has at relatively high
temperatures a density anomaly where the thermal expansion coefficient
changes sign.)

\begin{figure}
\centerline{
\psfig{figure=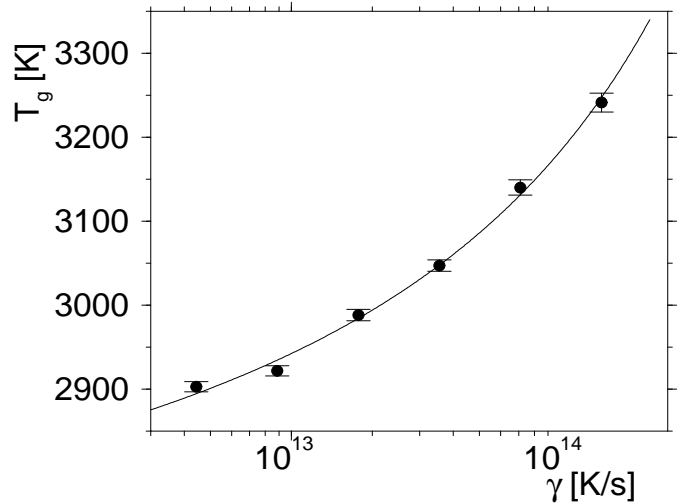,width=9.0cm,height=6.5cm}
}
\caption{\label{fig4}Effective glass transition temperature $T_g(\gamma)$
plotted vs. the cooling rate $\gamma$, for molecular dynamics simulations of SiO$_2$
using the BKS potential \cite{45} and estimating $T_g({\gamma})$ from
intersection points of fit functions to the enthalpy, as described in the
text. All data are based on averages over $10$ statistically independent
runs. The curve is a fit to the function given in Eq.~(\ref{eq5}) of
the text, resulting in $T_{VF}=2525 K$. From Vollmayr {\it et al.} \cite{46}.}

\end{figure}

A particular dramatic failure with extrapolations to lower values of
$\gamma$ was encountered in an attempt to determine the cooling rate
dependence of the glass transition temperature $T_g(\gamma)$. As is
done in many experimental studies, one can fit a smooth function of
temperature to the liquid branch of the enthalpy (where the melt has not
yet fallen out of equilibrium) and another smooth function to the glass
branch of the enthalpy, and estimate $T_g(\gamma)$ from the temperature
where these two branches intersect. Fig.~\ref{fig4} shows a plot of $T_g
(\gamma)$ versus $\gamma$ - note the logarithmic scale for $\gamma$! - for
the simulation of SiO$_2$. One sees, first of all, that there is a very
strong dependence of $T_g(\gamma)$ on $\gamma$, with $T_g(\gamma)\approx
4000$K for $\gamma=10^{15}$K/s, while $T_g(\gamma)$ has decreased down
to about $T_g(\gamma)\approx 2900$K for $\gamma=10^{13}$K/s. In this
range of cooling rates, the dependence of $T_g(\gamma)$ is not linear
in $\log(\gamma)$.  Nonlinear variations of $T_g(\gamma)$ that are
qualitatively similar to those of Fig.~\ref{fig4} have been reported in
the experimental literature, too \cite{47}, and are typically described by

\begin{equation}
T_g(\gamma)=T_{VF} - B/ [ \log (\gamma A) ]
\label{eq5}
\end{equation}

\noindent
This dependence can be justified by assuming that the fluid falls out
of its (metastable) equilibrium when the time constant of the cooling,
$\gamma^{-1}$, equals the structural relaxation time $\tau (T)$ at $T=T_g
(\gamma)$, and by using the Vogel-Fulcher law from Eq.~(\ref{eq1}) for
$\tau (T)$, $ \tau(T)= A \exp [B/(T-T_{VF})]$.  Obviously, Eq.~(\ref{eq5})
does provide a very good fit to the data of the SiO$_2$ simulation, but
the resulting $T_{VF}=2525$K is rather unreasonable: Remember that the
{\it experimental} glass transition temperature is $T_g \approx 1450$K,
the melting temperature of crystalline SiO$_2$ is around $2000$K, and
$T_{VF}$ should be significantly lower than $T_m$ and even somewhat
below $T_g$, cf. Fig.~\ref{fig2}. We emphasize here that the failure
of the simulation to predict $T_g(\gamma=1$K/s) is not primarily due
to the inaccuracy of the pair potential since, as will be explained in
detail below, a different analysis of SiO$_2$ simulation data yields
much more reasonable results. The failure implied by Fig.~\ref{fig4}
simply comes from the fact that the 10-1000 picosecond timescale that
is basically probed here is too many orders of magnitude off from the
time scale relevant for the glass transition and that therefore an
extrapolation of the results becomes a insecure undertaking.

A better way to study amorphous silica is to fix density at a reasonable
value, for instance the experimental value, and equilibrate the system at
a temperature which is as low as possible. Present day simulations can
propagate a system of around 8000 ions over a time span of around 20ns
which allows for a full equilibration at $T=2750$K~\cite{48}. Longer
time are accessible for smaller systems. However, it was found that if
one has fewer than $O(10^3)$ ions the results are plagued with finite
size effects~\cite{49}.  Simulating a large system over this time scale
are on the forefront of what is feasible today, and require the use
of multi-processor super computers such as CRAY-T3E, making use of a
parallelization of the force calculation \cite{48,49,50}.

\begin{figure}
\centerline{
\psfig{figure=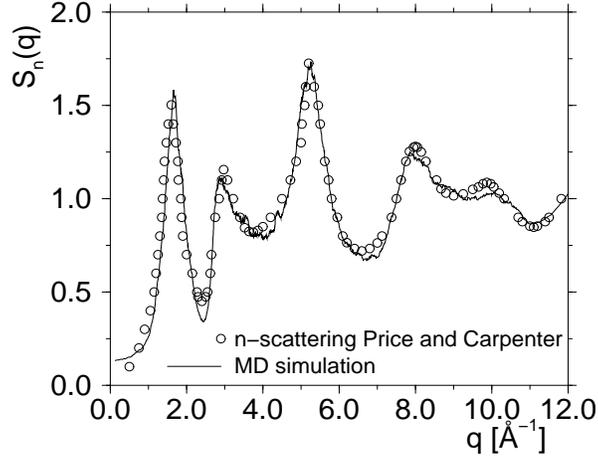,width=8.0cm,height=6.0cm}
}
\caption{\label{fig5}Static neutron structure factor of SiO$_2$
at room temperature $(T=300K)$ plotted versus wave-vector $q$. The full
curve is the molecular dynamics simulation of Ref. \cite{48}, using the
experimental neutron scattering lengths for Si and O atoms, while the symbols
are the neutron scattering data of Ref.~\cite{51}.  From Horbach
and Kob \cite{48}.}
\end{figure}

This well-equilibrated melt can then be used as a starting condition for a
cooling run at constant density. The advantage of this procedure is
that a state at $T=2750 K$ at the correct density is much closer in
local structure to the real glass, than the structures generated by
the procedures described above, and hence the spurious effects of the
by far too rapid quench are much less pronounced. This conclusion is
corroborated by a comparison of the simulated structure factor with
experiment \cite{51}, see Fig.~\ref{fig5}.  Given the fact that the
comparison in Fig.~\ref{fig5} does not involve any adjustable parameter
whatsoever, the agreement between simulation and experiment is quite
remarkable, and this reiterates our above conclusion that the potential
used \{Eq.~(\ref{eq4})\} is accurate enough, and should not be blamed
for discrepancies as discussed in connection with Fig.~\ref{fig4}.

\begin{figure}
\centerline{
\psfig{figure=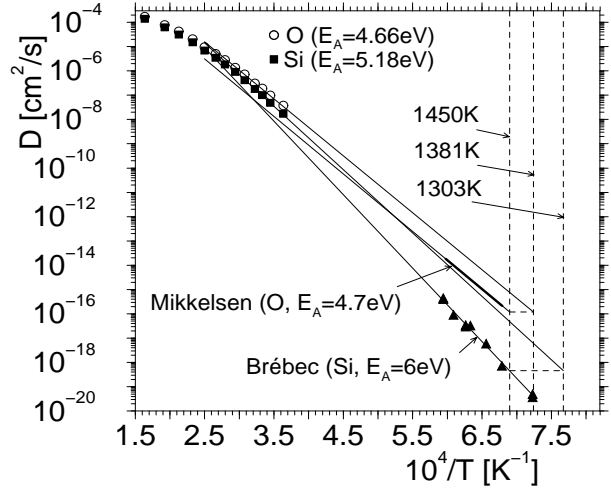,width=8.0cm,height=6.3cm}
}
\caption{\label{fig6}Plot of the self-diffusion constant $D$ of silicon
atoms (Si) and oxygen atoms (O) in molten SiO$_2$ as a function of
inverse temperature. The symbols in the upper left part are the results
from molecular dynamics simulations and the data in the lower right part
stems from experiments~\cite{52,53}. The thin straight lines show simple Arrhenius
behavior $\{D \propto \exp (-E_A/k_BT)\}$ with various choices of the
activation energy $E_A$, as indicated in the figure. The vertical broken
lines indicate the experimental glass transition temperature, $T_g=1450$K,
as well as values for $T_g$ that one obtains if one extrapolates the data
from the simulations to low temperatures and then estimates $T_g$ from the
experimental value of the O diffusion constant $\{D_{\rm O}(T=T_g^{{\rm
sim}})=10^{-16}$cm$^2$/s $\Rightarrow T_g^{{\rm sim}}=1381$K or the Si
diffusion constant, respectively $\{D_{\rm Si} (T=T_g^{{\rm sim}})=5\cdot
10^{-19}$cm$^2$/s $\Rightarrow T_g ^{{\rm sim}} = 1303$K. From Horbach
and Kob \cite{48}.}
\end{figure}

For the temperatures at which one can equilibrate the system,
i.e. here 2750K and higher, it is also possible to determine the
self-diffusion constants of Si and O atoms from the simulation. This
is done by calculating the mean square displacements $\langle
|\vec{r}_i(t)-\vec{r}_i(0)|^2 \rangle=\Delta r_{\alpha}^2 (t)$ of the
particles of type $\alpha \in \{\mbox{Si,O}\}$, and apply the Einstein
relation $\Delta_{\alpha}(t)=6D_\alpha t$ in the regime of late times
where the dependence of $\Delta_{\alpha}(t)$ on $t$ is in fact linear
\cite{48,49,50}\}.  The result is shown in Fig.~\ref{fig6}, where also
the respective experimental data \cite{52,53} are included. As one
can see from Fig.~\ref{fig6}, one needs to cover $16$ decades, from
$10^{-4}$cm$^2$/s to  $10^{-20}$cm$^2$/s, to cover the full range including
simulation results and experiments, but the simulation results alone
are actually restricted to the first four decades of this range only. The
straight lines fitted on this Arrhenius plot to the experiment as well as
to the simulation show that in this case a bold extrapolation actually
is rather successful - but of course there is no guarantee that this will work
similarly well in other cases.

A very interesting aspect of the temperature dependence of the diffusion
constants is that there are strong deviations from Arrhenius behavior
at very high temperatures. It turns out that this region is rather
well described by a power law, $D \propto (T-T_c)^\gamma$, as it is
implied by mode coupling theory, see Eq.~(\ref{eq3}) with $D \propto
\tau^{-1}$. In fact, this conclusion is strongly corroborated by a
detailed analysis of the intermediate scattering function $\phi_q (t)$
for wave-vector $q$ and various other quantities~\cite{48}. This finding
is somewhat surprising, however, since $T_c \approx 3330 K $ \cite{48},
i.e. far above the melting temperature of crystalline SiO$_2$! Thus it
is no surprise that experimental results had not given hint that mode
coupling theory also describes a ``strong'' glassformer such as SiO$_2$
(where $\tau$ and $\eta(T)$ follow a simple Arrhenius behavior over
a wide range of temperature). Nevertheless, this discovery that a
critical temperature exists also for SiO$_2$ is of great interest,
because it suggests that the differences of the relaxation dynamics
between different glassforming fluids are of a quantitative nature only,
while qualitatively the behavior is always the same.

\section{Parallel tempering}

One of the major reasons for the slowing down of the dynamics of
(atomistic) glass forming systems is that at low temperatures each atom is
trapped in a cage formed by its surrounding neighbors. On the other hand
the atom itself is part of a cage that trap the neighboring atoms. With
decreasing temperature each of these cages becomes stiffer and stiffer and
finally each atom can perform only a rattling motion, i.e. the system has
become a fluid that doesn't flow anymore, i.e. a glass. The basic idea of
the parallel tempering method is to help the particles to escape their
local cage by supplying them with sufficient kinetic energy to overcome
the local barrier. Originally proposed for spin models~\cite{21a,21},
the method has been found to be also useful for off-lattice systems. A
recent review on the method can be found in Refs.~\cite{21b,21c}. In
the following we discuss briefly how the method is implemented in practice.

If we denote the Hamiltonian of the system as $H=K({\bf p})+E({\bf q})$,
where $K$ and $E$ are the kinetic and potential energy, respectively,
and ${\bf p}=(p_1,p_2,\ldots,p_N)$ and ${\bf q}=(q_1,q_2,\ldots,q_N)$
are the momenta and coordinates of the particles, we construct a new
Hamiltonian $\cal{H}$ as follows:

Make $M$ independent copies of the Hamiltonian $H$: $H_i=K({\bf
p}_i)+E({\bf q}_i)$. Here the $\bf{p}_i$ and $\bf{q}_i$ are the momenta
and coordinates belonging to the $i-$th subsystem. $\cal{H}$ is then
defined as

\begin{equation}
{\cal H}({\bf p}_1, \ldots, {\bf p}_M, {\bf q}_1, \ldots {\bf q}_M) = \sum_{i=1}^M
H_i({\bf p}_i,{\bf q}_i)=\sum_{i=1}^M K({\bf p}_i)+\Lambda_i E({\bf q}_i).
\label{pt1}
\end{equation}

\noindent
The $1=\Lambda_1>\Lambda_2 > \ldots \Lambda_M$ are constants which
we will use later.  We now make a molecular dynamics simulation of the
Hamiltonian $\cal{H}$ at a constant temperature $T=\beta_0^{-1}$. After
a certain time interval $\Delta t_{\rm PT}$ we attempt to exchange the
two configurations $m$ and $n$ belonging to two neighboring systems
(i.e. $m=n\pm 1$). Whether or not the swap of these two configurations
is accepted depends on a Metropolis criterion with a acceptance probability

\begin{equation}
w_{m,n}= \left\{
\begin{array}{ll}
1,&\qquad \Delta_{m,n}\le 0\\
\exp(-\Delta_{m,n}),&\qquad\Delta_{m,n}> 0,
\end{array}
\right.
\label{pt2}
\end{equation}
where $\Delta_{m,n}=\beta_0(\Lambda_n-\Lambda_m)(E({\bf q}_m)-E({\bf
q}_n))$. Since the normal molecular dynamics simulation as well as
the Monte Carlo procedure on time scale $\Delta t_{\rm PT}$ fulfill
the condition of detailed balance, the whole algorithm does so also,
i.e. after a sufficiently long time the system composed by the subsystem
will converge to a Boltzmann distribution. Note that in the systems
with a small value of $\Lambda$ the interaction between the particles
is weakened (see Eq.~(\ref{pt1})). Therefore it can be expected that
the particles in these systems move faster than those in systems with
a large value of $\Lambda$. Another way to see this is to say that each
system is simulated at a different temperature and that periodically the
temperature of the system is increased or decreased (hence the name of
the algorithm). This walk in temperature space should thus allow the
system to overcome the local barriers formed by the above mentioned
cages and thus to propagate faster in configuration space.

\begin{figure}
\centerline{
\psfig{figure=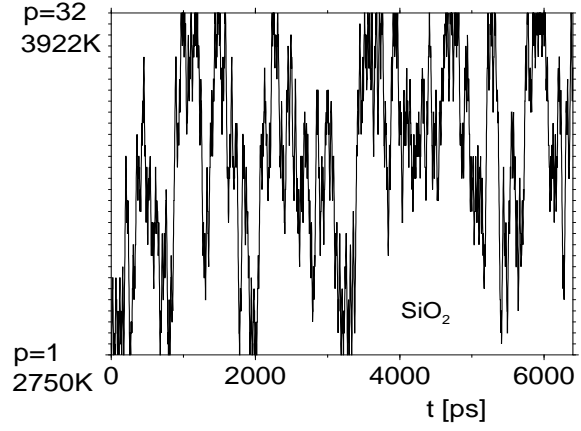,width=7.5cm,height=5.7cm}
}
\caption{Time dependence of the coupling constant for a parallel
tempering simulation of liquid SiO$_2$. The number of particles was
336 and the number of subsystems was 32. Note that the shown subsystem
visits all the different coupling constants several times, thus giving
evidence that the overall system has indeed reached equilibrium. From
Kob {\it et al.}~\cite{22}.
\label{fig7}}

\end{figure}

Note that this algorithm has a substantial number of parameters, all
of which influence its efficiency considerably. In order that the
acceptance probabilities of Eq.~(\ref{pt2}) are reasonably high, the
coupling constants $\Lambda_i$ should not be too different. On the other
hand one wants that $\Lambda_M$ is as small as possible since this will
lead to a fast propagation of the system at this temperature. Therefore
one is forced to choose a relatively large value of $M$. This in turn
is, however, not good for the overall performance of the algorithm
since in order to be ergodic each configuration has to make a random
walk in $\Lambda-$space, and the time to do this increases like
$M^2$. Last not least there is the exchange time $\Delta t_{\rm PT}$
which should not be too small since then the system just swaps back and
forth configurations that are very similar. On the other hand $\Delta
t_{\rm PT}$ should also not be too large, since one needs these type
of moves in order to explore the $\Lambda-$space quickly. The optimal
choice of these parameters is currently not known and still the focus
of research~\cite{stuehn00}. A further problem is to find out after
which time the system $\cal{H}$ has really equilibrated. It seems that to
guarantee this it is not sufficient that every subsystem has visited every
point in $\Lambda-$space~\cite{stuehn00,demichele01}. A good random walk
should look like the one shown in Fig.~\ref{fig7}. Furthermore we point
out that it might be possible that a suboptimal choice of these parameters
might make the whole algorithm rather inefficient~\cite{demichele01}.

\begin{figure}
\centerline{
\psfig{figure=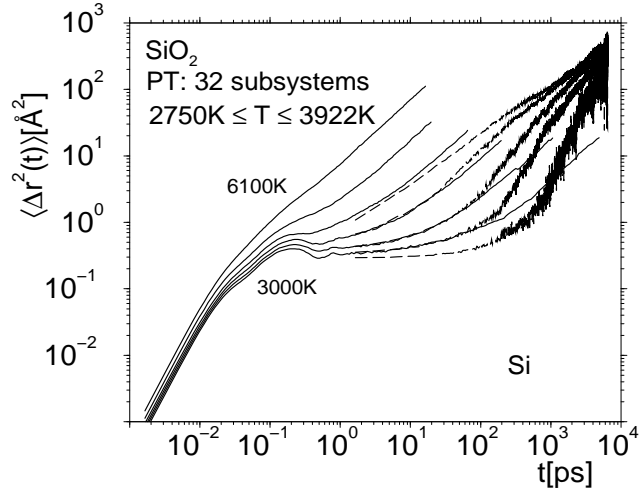,width=8.5cm,height=6.3cm}
}
\caption{Time dependence of the mean squared displacement of Si in SiO$_2$
at different temperatures. The dashed lines are from parallel tempering
runs and correspond to temperatures 3922K, 3585K, 3235K, 3019K, and 2750K
(top to bottom). The solid lines are from conventional molecular dynamics
runs and correspond to temperatures 6100K, 4700K, 4000K, 3580K, 3250K,
and 3000K (top to bottom). From Kob {\it et al.}~\cite{22}.
\label{fig8}}
\vspace*{-5mm}

\end{figure}

If the above mentioned parameters of the algorithm are chosen well,
the parallel tempering method can indeed speed up the equilibration
of the system considerably. This is demonstrated in Fig.~\ref{fig8}
where we show the mean squared displacement of the silicon atoms in
SiO$_2$ as a function of time. From the figure we see that at the lowest
temperatures the mean square displacement increases by about a factor of
100 faster than the corresponding curve obtained from the conventional
molecular dynamics simulation. From the figure it becomes also clear
that the parallel tempering slows {\it down} the dynamics of the system
at high temperatures. This is due to the fact that these systems are
coupled to the ones at the low temperatures and hence cannot propagate
as fast anymore.

Before we conclude this section we mention that the parallel tempering
algorithm has been found to be also very efficient for the equilibration
of the Potts glass discussed in the next section. Thus, although the
algorithm might have some problems for certain systems or values of
parameters, there are models where it seems to work very well.

\section{An abstract model for static and dynamic glass transitions:
The $10$-state mean field Potts glass}

In this section we are concerned with a model for which it is known
exactly that there is a {\it dynamical} (ergodic to nonergodic) transition
at a temperature $T_D$ and a second, {\it static}, transition at a lower
temperature $T_0<T_D$, where a static glass order parameter $q$ appears
discontinuously: the infinite range $p$-state Potts glass with $p>4$
\cite{54,55,56,57,58,59,60}. In this model, one has Potts ``spin''
variables $\sigma_i$ which can take one out of $p$ discrete values
which we simply label from 1 to $p$, $\sigma_i \in \{1,2,\ldots ,p\}$,
where $i$ labels the ``sites'', $i=1,2, \ldots ,N$. An energy $p J_{ij}$
is gained if two spins $\sigma _i, \sigma _j$ are in the same state,

\begin{equation} \label{eq6}
{\mathcal{H}} =-\sum _{i<j} J_{ij} (p \delta _{\sigma _i \sigma_j}-1) \; .
\end{equation}

Every spin interacts with every other spin via an interaction $J_{ij}$ which is Gaussian
distributed, i.e.

\begin{equation}
P(J_{ij})=\left [\sqrt{2 \pi} (\Delta J) \right ]^{-1} 
\exp \{{-(J_{ij}-J_{0})^2 /[2(\Delta J)^2]}\} \;.
\label{eq7}
\end{equation}

\begin{figure}
\centerline{
\psfig{figure=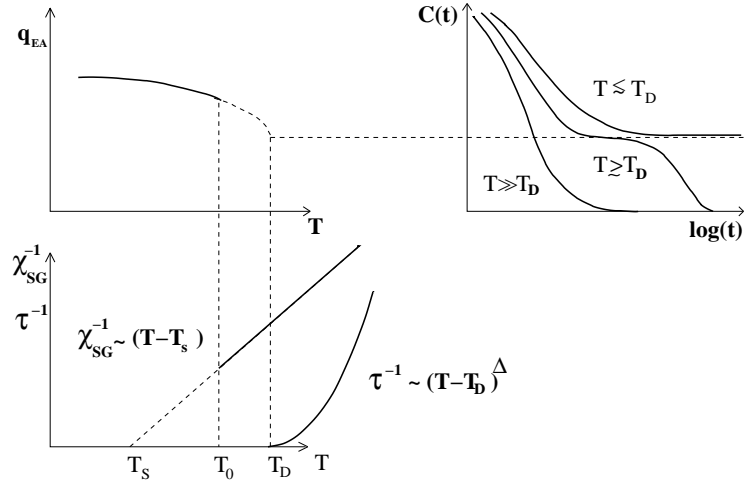,width=10.0cm,height=6.6cm}
}
\caption{\label{fig9}Mean-field predictions for the $p$-state Potts glass
with $p>4$. The spin glass order parameter, $q_{\rm EA}$, is nonzero only
for $T<T_0$ and jumps to zero discontinuously at $T=T_0$. The spin glass
susceptibility $\chi_{\rm SG}$ follows a Curie-Weiss-type relation with
an apparent divergence at $T_{\rm S}<T_0$. The relaxation time $\tau$
diverges already at the dynamical transition temperature $T_D$. This
divergence is due to the occurrence of a long-lived plateau of height
$q_{\rm EA}$ in the time-dependent spin autocorrelation function
$C(t)$. The discontinuous transition of the order parameter, however,
is not accompanied by a latent heat. Therefore, there is no jump in
the entropy at $T_0$, but only a kink occurs. The extrapolation of the
high-temperature branch of the entropy would vanish at a ``Kauzmann
temperature'' $T_K=[(1/4) (p-1) / \ln p] ^{1/2} T_{\rm S}\approx 0.988
T_{\rm S}$. From Brangian {\it et al.} \cite{26}.}
\end{figure}

\noindent
Here the mean $J_0$ and the width $\Delta J$ are normalized such that

\begin{equation} 
J_0 \equiv [J_{ij}]_{av} =\widetilde{J}_0 /(N-1), \quad (\Delta J)^2 
\equiv [J_{ij}^2]_{av} -[J_{ij}]^2_{av} = \Delta \widetilde{J}/(N-1) \;,
\label{eq8}
\end{equation}

\noindent
a choice that ensures a sensible thermodynamic limit. We fix the
temperature scale by choosing $\Delta \widetilde{J} \equiv 1$, and set
the mean of the distribution ``antiferromagnetic'', $\widetilde{J} =3-p$,
in order to avoid any tendency towards ferromagnetic order. (Note that for
$p=2$ this model would reduce to the standard Ising mean field spin glass
(Sherrington-Kirkpatrick model) \cite{61}, but we shall be concerned
with $p=10$ here.) This model, which due to the choice Eq.~(\ref{eq7})
exhibits quenched random disorder already in the high temperature phase
above the glass transition, can be solved exactly in the thermodynamic
limit \cite{54,55,56,57,58,59,60}. One finds (Fig.~\ref{fig9}) that
slightly above $T_D$ the dynamic auto-correlation function of the spins
exhibits a two-step decay, in that a plateau develops whose life-time
diverges at $T_D$. It is important to note that this behavior is
described {\it exactly} by mode coupling equations of the same type as
they occur for the structural glass transition \cite{3}! This shows that
this rather abstract model might be more similar to a real structural
glass than one would expect at a first glance. At a lower temperature
$T_0$, a static glass transition occurs \cite{54,55,56,57,58,59,60},
where a static order parameter appears discontinuously. Interestingly
the static response function does not diverge at $T_0$, i.e. the glass
susceptibility is still finite here. The entropy does not have a jump at
$T_0$, but shows only a kink. Thus there is no latent heat associated with
this transition! A Kauzmann temperature $T_K$, where the (extrapolated)
entropy of the high temperature phase would vanish, also exists, but in
this case clearly $T_K < T_0$ and $T_K$ does not have a physical meaning.

\begin{figure}
\centerline{
\psfig{figure=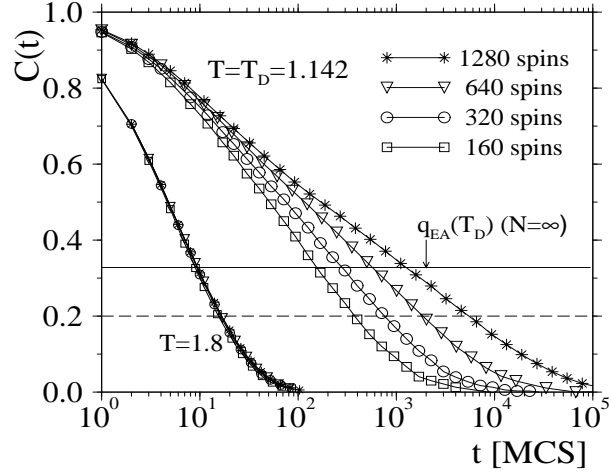,width=8.0cm,height=6.5cm}
}
\caption{\label{fig10}Time dependence of the autocorrelation function
$C(t)$ of the spins in the 10-state mean field Potts glass. $C(t)$ is normalized
such that $C(t=0)=1$ and $C(t \rightarrow \infty)=0$ for $T>T_D$. Time is
measured in units of Monte Carlo steps per spin [MCS].  Two temperatures
are shown, $T=1.8$ and $T=T_D=1.142$ \cite{60}, for several values of
$N$. The solid horizontal line indicates the theoretical value of the
Edwards-Anderson order parameter $q_{\rm EA} (T_D)\equiv C(t \rightarrow
\infty)$ at $T \rightarrow T_D^{-}$ for $N\to \infty$\cite{60}. The
horizontal dashed line shows the value used to define the relaxation time
$\tau$, $C(t \equiv \tau)=0.2$. From Brangian {\it et al.} \cite{24}.}
\end{figure}

It is of course interesting to know if computer simulations can identify
the static and dynamic glass transition in a model for which one knows
from the exact solutions \cite{54,55,56,57,58,59,60} that all these
glass transitions do indeed exist. Surprisingly, the answer to this
question is ``no'' since very strong finite size effects are present. In
particular it is even hard to see that the above mentioned plateau in
the autocorrelation function develops as one approaches the temperature
$T_D$ of the dynamical transition (Fig.~\ref{fig10}) \cite{24}. This is
demonstrated in Fig.~\ref{fig10} where we show the autocorrelation function
of the Potts spins as a function of Monte Carlo time, for $160 \leq N
\leq 1280$. Note that this range is of the same order of magnitude as the
particle numbers used for simulations of the structural glass transition,
using models such as the binary Lennard-Jones fluid \cite{62} or similar
models. No evidence for strong finite size effects was ever found for
the latter models if $N$ was larger than $\approx 1000$~\cite{kim00}.
Thus, {\it a priori} it is not at all obvious that system sizes of the
order $10^3$ are completely insufficient to characterize the dynamics
of a system in the thermodynamic limit. However, from Fig.~\ref{fig10}
we must conclude that for the present system this is indeed the case,
at least for temperatures close to the dynamical transition temperature
$T_D$. This is in contrast with the behavior at a high temperature,
e.g. $T=1.8$. From the figure we recognize that at this temperature
there are hardly any finite size effects and that the data have nicely
converged to the thermodynamic limit even for modest system sizes.

\begin{figure}
\centerline{
\psfig{figure=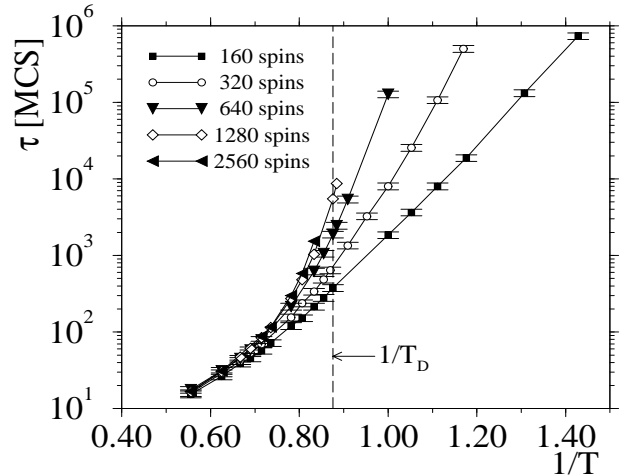,width=8.0cm,height=6.5cm}
}
\caption{\label{fig11}Arrhenius plot of the relaxation time $\tau$ of the
10-state mean field Potts glass model for different system sizes. Error
bars of $\tau$ are mostly due to sample-to-sample fluctuations. The
vertical dashed line is the location of $T_D$ where, for $N\to
\infty$, the relaxation time is predicted to diverge with a power-law.
From Brangian {\it et al.} \cite{24}.}

\end{figure}

Brangian {\it et al.} \cite{23,24,25,26} defined a relaxation time $\tau$
from the time $t$ that it takes the autocorrelation function to decay to
the value $C(t=\tau)=0.2$ (broken straight line in Fig.~\ref{fig10}). This
time is plotted logarithmically versus $1/T$ in Fig.~\ref{fig11}, so
an Arrhenius behavior would be a straight line on this plot. One can
see rather clearly a crossover from a power law divergence (that would
emerge fully in the limit $N \rightarrow \infty$ for $T>T_D$) to the
Arrhenius law at low $T$. This behavior qualitatively resembles the
behavior expected for structural glasses where the different valleys in
the rugged energy landscape for $T<T_0$ are separated by finite (free)
energy barriers. In contrast to this one knows that in the Potts glass
in the limit $N \rightarrow \infty$ these barriers are truly infinite
if $T<T_D$, and hence the dynamics is strictly nonergodic.

\begin{figure}
\centerline{
\psfig{figure=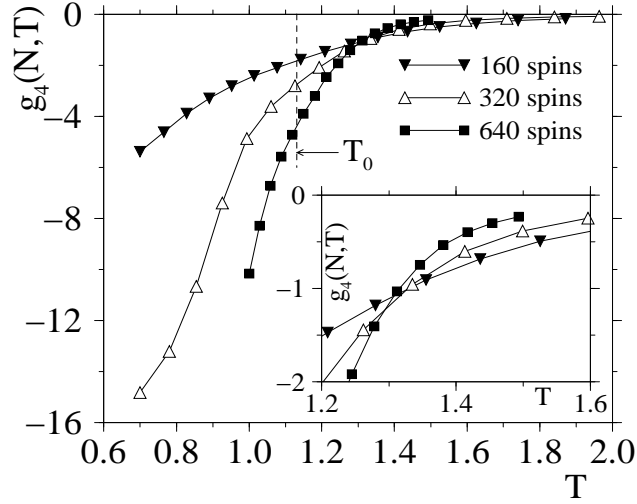,width=8.3cm,height=7.0cm}
}
\vspace*{-5mm}

\caption{\label{fig12}Temperature dependence of the order parameter
cumulant $g_4(N,T)=\frac{(p-1)^2}{2} \Big\{1 + \frac{2} {(p-1)^2}
-\frac{[\langle q^4 \rangle]_{av}} {[\langle q^2 \rangle]_{av}} \Big\}
$ for the 10-state mean field Potts glass for three choices of $N$
$(N=160$, 320$ \, $and$ \, $640$)$. (Here $q$ is the order parameter.)
The vertical straight line shows the location of the static transition
temperature $T_0$ as predicted by the exact solution \cite{60}. The inset
is an enlargement of the region where the three curves intersect. From
Brangian {\it et al.} \cite{24}.}

\end{figure}

Similar finite size effects affect also the behavior of static properties
\cite{23,24,25,26}. One might wonder whether it is possible to use these
finite size effects to apply standard finite size scaling analyses to
extract reliable information on the location of the static transition
temperature from the simulations. Unfortunately the answer is ``no'': As
figure ~\ref{fig12} shows, the standard method \cite{63} of locating
a static transition from the intersection point of the order parameter
cumulant gives rather misleading results here since the curve seem(!) to
intersect at a wrong temperature. Thus one must conclude that there is
need to better understand finite size effects for such unconventional
glass transitions as sketched in Fig.~\ref{fig9}, before one can study
them reliably with simulations.

\section{The bead-spring model: A coarse-grained model for the study of
the glass transition of polymer melts}

We now draw attention to a model which is intermediate between the
abstract model as considered in the previous section and the chemically
realistic model of silica melts discussed in Sec. 2. This intermediate
model is a coarse-grained model of glassforming polymer melts. Short
polymer chains are described by a bead-spring model, with a chain length
of $N=10$. The (effective) monomers interact with each other 
via a truncated and shifted Lennard-Jones potential,

\begin{equation} \label{eq9}
U_{LJ} (r)=4 \varepsilon [(\sigma /r)^{12} - (\sigma /r)^6] + C 
\, ,  \quad r \leq r_c=2.2^{1/6} \sigma
\end{equation}

\noindent
while $U_{LJ}(r)=0$ if  $r>r_c$. The constant $C$ is chosen such that
$U_{LJ}$ is continuous at $r=r_c$.

The spring potential present between two neighboring beads is given by

\begin{equation} \label{eq10}
U_{\rm FENE} (l)=-(k/2) R_0^2 \log [1-(l/R_0)^2] \,
\end{equation}

with the following values of the constants \cite{27}:

\begin{equation} \label{eq11}
\varepsilon=1 \, , \sigma=1 \, , k=30 \, , R_0=1.5 \, .
\end{equation}

\noindent
This choice for the parameters creates frustration in the model:
the minimum of the bond potential along the chain occurs at a position
$l_{\rm min} \approx 0.97$ that is incompatible with the minimum position
$r_{\rm min} \approx 1.13$ of the Lennard-Jones potential, as far as
the formation of simple crystal structures is concerned. This conflict
between these two length scales prevents crystallization very efficiently,
and the resulting structure of the melt and the corresponding glass
resembles corresponding experimental data very nicely, as has already
been demonstrated in Fig.~\ref{fig1}.

If one carries out ``slow'' cooling experiments one finds that
the volume per monomer shows at a temperature $T_g \approx 0.41$
a kink~\cite{28}. This signals that the system has changed from
the liquid branch to the glass branch and hence has fallen out of
equilibrium. Qualitatively, the data looks again very similar to that of
corresponding experiments \cite{64}. However, if one compares experiment
and simulation more quantitatively, one notes again a big disparity in the
cooling rates: In the simulation the temperature was reduced by $\Delta
T=0.02$ every $500 000$ MD time steps, each time step being $\delta t=
0.002 \tau_{MD}$ with $\tau_{MD}=\sigma (m/ \varepsilon)^{1/2} \,$,
$m$ being the effective mass of the monomeric units. If one estimates
that $\tau_{MD}$ corresponds roughly to $10^{-11}$s, and that $T=1$
corresponds to $500$K, one arrives at a cooling rate of $\Delta T/
\Delta t\approx 10^9$K/s. While this estimate is three orders of
magnitude smaller than the corresponding cooling rate for the silica
melts \cite{46}, it is still many orders of magnitude larger than the
corresponding experimental cooling rates. Hence also in this case there
is a huge gap between the cooling rates accessible in simulations and
those used in real experiments.

\begin{figure}
\centerline{
\psfig{figure=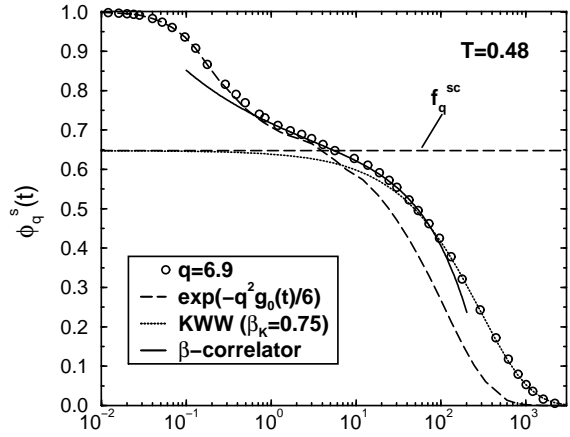,width=8.3cm,height=7.0cm}
}
\vspace*{-8mm}

\caption{\label{fig13}Comparison of the incoherent intermediate scattering
function $\phi^{s}_{q} (t)$ for the bead-spring model at $T=0.48$
and $q \approx 6.9$ [$\approx$ maximum of $S(q)$, cf. Fig.~\ref{fig1}]
with various approximations: a Gaussian approximation (dashed line),
$\phi_q^S = \exp [-q^2 g_0 (t)/6]$, where $g_0(t)$ is the mean square
displacement of the monomers. The mode coupling fit for the regime of
the so-called ``$\beta-$relaxation'' (solid-line) and a fit with
the Kohlrausch function \{Eq.~(\ref{eq2}), dotted line\} also are
included. The non-ergodicity parameter $f$ is indicated as a horizontal
dashed line. From Baschnagel {\it et al.}  \cite{13}.}

\end{figure}

This model yields also qualitatively very reasonable results for the
relaxation dynamics: The self-diffusion constant can be fitted well by the
Vogel-Fulcher law given by Eq.~(\ref{eq1}), with $T_{VF} \approx 0.34$,
below the kink temperature $T_g \approx 0.41$. The mode coupling critical
temperature is located at $T_c \approx 0.45$, above the kink temperature,
and the ratios $T_c/T_g$ and $T_c/T_{VF}$ are quite reasonable. Although
in the simulation only $1200$ monomers were used, a nice plateau is
found in the intermediate incoherent scattering function $\phi_q^s (t)$,
see Fig.~\ref{fig13}. Hence one can conclude that no strong finite size
effects are present for this model.

\begin{figure}
\centerline{
\psfig{figure=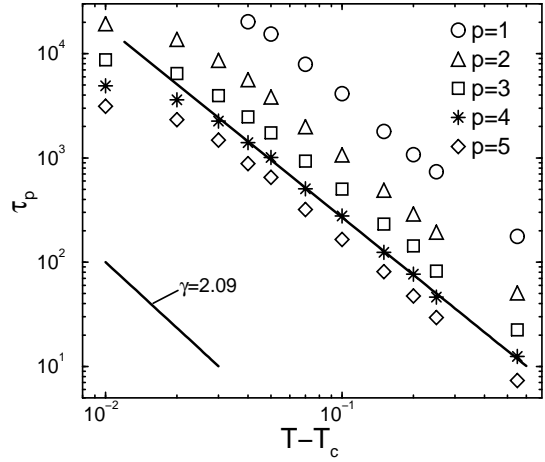,width=8.2cm,height=7.0cm}
}
\vspace*{-3mm}

\caption{\label{fig14}Variation of the relaxation time $\tau_p$ of the
Rouse modes with the mode index $p$ for the bead spring model plotted
vs. $T-T_c$, showing also a power law fit for $p=4 (\gamma_p = 1.83 \pm
0.02$). Within the error bars, this slope provides a reasonable
fit for all $p$ shown. From Baschnagel {\it et al.} \cite{29}.}

\end{figure}

Also the Rouse modes \cite{65} which describe the mesoscopic Brownian
motion of the polymer chains on length scales that are between
monomer-monomer distances and the coil size, are found to relax over
almost two decades in $T-T_c$ with relaxation times that show the mode
coupling power law \cite{29}, see Fig.~\ref{fig14}. Only very close
to $T_c$, for $T \leq 0.46$, can one see small indications that the
singularity at $T_c$ is in fact rounded off. This model has allowed many
very impressive tests \cite{31,33} of mode coupling theory, similar to
an often studied binary Lennard-Jones mixture \cite{62,nauroth97}. But
similar to the case in the latter model, it has so far turned out
impossible to study temperatures for $T<T_c$ in thermal equilibrium. And
none of these models - neither the model for SiO$_2$, nor the binary
Lennard-Jones model \cite{62} nor the present beadspring model - could
provide any clarification about the validity of the entropy theory
\cite{16}.

\section{The bond fluctuation model approach to glassforming polymer melts}

\begin{figure}
\centerline{
\psfig{figure=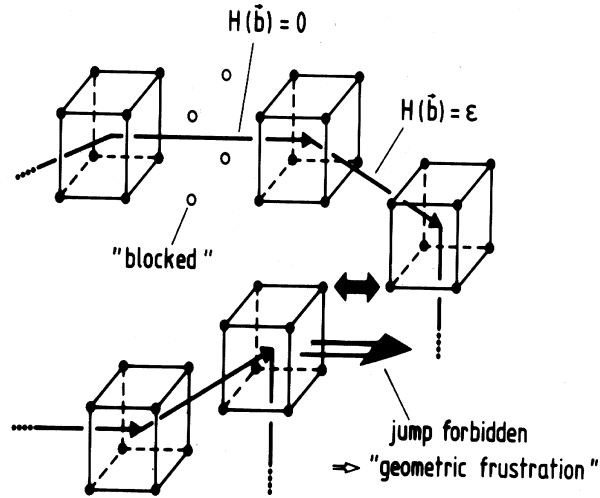,width=8.2cm,height=7.0cm}
}
\vspace*{-3mm}

\caption{\label{fig15}Sketch of a possible configuration of monomers
belonging to two different chains in the bond fluctuation model of a
polymer melt. For one monomer of the lower chain, an attempted move is
indicated; this jump is forbidden, however, since it violates the excluded
volume constraint. Also the choice of a two-state energy function is
indicated, namely $\mathcal{H}(\vec{b})=0$ if the bondvector $\vec{b}$
equals $\vec{b}_{\rm min}=(0, \, 0, \,  \pm \, 3a)$ or a permutation
thereof ($a$ is the lattice spacing, chosen as unit of length in the
following), and $\mathcal{H}(\vec{b})=\varepsilon=1$ else. Note that if a
bond takes a ground state bond $\vec{b}_{\rm min}$ it blocks automatically
4 sites (the 4 sites are highlighted by empty circles). From Baschnagel
{\it et al.} \cite{35}.}

\end{figure}

The bond fluctuation model \cite{6,34,35,36,37,38,39,40,41,42,43}
is an even more abstract model of polymers than the bead-spring model
discussed in the previous section, since it forces the chains to ``live''
on a simple cubic lattice, and all motions on scales smaller than a
lattice constant are completely suppressed. In this model a polymer
is represented again as a chain of effective monomers connected by
effective bonds, but now each effective monomer is described by an
elementary cube on the lattice that blocks all 8 sites at the corners
of the cube from further occupation (Fig.~\ref{fig15}). The length of
the effective bonds is allowed to vary from 2 to $\sqrt{10}$ lattice
constants (taken as length unit in this section). The only nonbonded
interaction is the one of excluded volume. The dynamics of the random
conformational changes of the real polymer is represented in a crude
way by attempted hops of randomly chosen monomers in randomly chosen
lattice directions. If about one half of all lattice sites are occupied,
the system behaves like a dense melt, and even short chains with chain
length $N=10$ show already typical polymer-properties, e.g. the scaling
of the radius of gyration with $\sqrt N$, etc.

Since real polymers show with decreasing temperature an increase of
the persistence length and hence of the chain radius, it is natural
to model this effect by an effective potential $U(l)$ for the length
of the effective bonds, energetically favoring long bonds. If one
chooses as a minimum of this potential $U(l_{\rm min}=3)=0$ while
$U(l)=\varepsilon=1$ for all other bond lengths  $l$, one also
incorporates ``geometric frustration'' (Fig.~\ref{fig15}) into the
model: Each bond that reaches its ground state wastes the four lattice
sites in between the adjoining effective monomers, which are completely
blocked for further occupation. From the point of view of packing as many
effective monomers as possible in a dense melt on the lattice, the bonds
that waste lattice sites are very unfavorable. Hence configurational
entropy favors short bonds that do not waste any other lattice sites
for further occupation. Thus a conflict between entropy and energy is
created, which is responsible for the glass transition observed in the
Monte Carlo simulations of this model.

This model has the big technical advantage that it can be equilibrated
even at relatively low temperatures by the so-called ``slithering snake
algorithm''. In this type of Monte Carlo moves one randomly attempts to
remove a bond from one chain end and attach it to the other chain end
in a randomly chosen orientation \cite{40}. Although this algorithm
does not correspond to any physically realistic dynamics of polymers
it is a perfectly admissible Monte Carlo move for studying equilibrium
properties. Using this algorithm, thermal equilibrium can be established
at rather low temperatures, such as $T=0.16$, where after $10^7$ steps
with the conventional ``random hopping'' algorithm the autocorrelation
of the end-to-end vector of the chains still has not decayed below
90\% of its starting value~\cite{42}.  If we wish to study dynamical
properties of this model, we first perform a run with this slithering
snake algorithm, to obtain initial states that are characteristic for
thermal equilibrium. Subsequently we can start a run with the normal
random hopping moves of the effective monomers, which thus yields
a physically reasonable description of the dynamics \cite{42}. If
one estimates that one Monte Carlo Step per monomer corresponds to
about $10^{-12}$ seconds in real time, a run of $10^7$ steps would
reach a physical time of $10^{-5}$ seconds, which is several orders of
magnitude longer than the typical time scales accessible with molecular
dynamics. Using this algorithm it was hence possible to make a very nice
test of mode coupling theory \cite{38,39}, resulting in $T_c \approx 0.15$
while \cite{42} $T_{VF}=0.125 \pm 0.005$. However, the investigation
of the relaxation dynamics in the regime $T_{VF}<T \leq T_c$ seems to
be very difficult also in the framework of this lattice model, and in
fact has not yet been attempted.

Using the bond fluctuation model it was also determined how the glass
transition temperature $T_g$ depends on the length of the chain~\cite{37}
and the results are compatible with the law

\begin{equation} \label{eq12}
T_g(\infty) - T_g(N) \propto 1/N \,.
\end{equation}

\begin{figure}
\centerline{
\psfig{figure=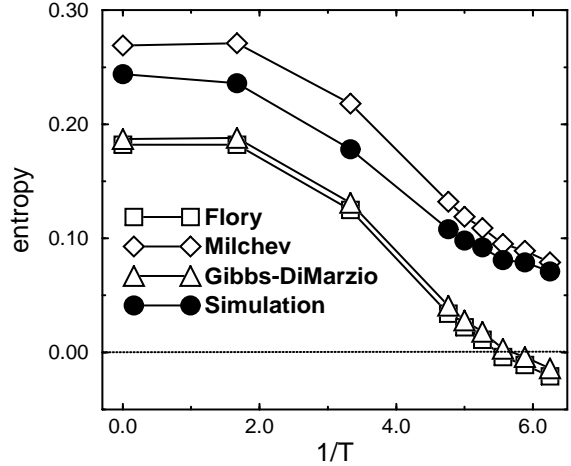,width=8.2cm,height=7.0cm}
}
\vspace*{-5mm}
\caption{\label{fig16}Comparison of the temperature dependence of the
entropy per lattice site as obtained from the simulation of the bond
fluctuation model (open circles) with the theoretical predictions of Gibbs
and Di Marzio \cite{16}, Flory \cite{67} and Milchev \cite{68}. Note that
the estimates for $T_c$ and $T_{VF}$ are $T_c \approx 0.15$ and $T_{VF}
\approx 0.125$. Therefore the vanishing of the entropy at $T \approx 0.18$
is an artifact due to inaccurate approximations involved
in the calculation of $S(T)$ via the entropy theory \cite{16}. From
Wolfgardt {\it et al.} \cite{41}.}
\end{figure}

Such a dependence has also been found experimentally \cite{66}, and is
one of the most notable predictions of the entropy theory of Gibbs and
Di Marzio \cite{16}. Therefore many experimentalists believe that this
theory is correct. However, this conclusion is premature, as a study
of the configurational entropy for the present lattice model shows
(Fig.~\ref{fig16}). While the entropy does indeed decrease rather
strongly with increasing value of inverse temperature, starting out
from an ``athermal melt'' (corresponding to infinite temperature), this
decrease becomes slower when one approaches the vicinity of $T_c$, and
the simulation data do not show that the entropy vanishes, although they
also cannot rule it out that this happens at a $T$ far below $T_c$. However,
if one works out the Gibbs-Di Marzio theory \cite{16} explicitly for the
present lattice model (all the input parameters of the theory \cite{16}
can also be extracted from the simulation, so there are no adjustable
parameters whatsoever in this comparison!), one sees that the theory
underestimates the actual entropy considerably at all temperatures.
In particular this failure is responsible for the vanishing of the
entropy at $T_K \approx 0.18$, which obviously is a spurious result,
since this temperature is even higher than $T_c$, well in the melt
regime where the polymer system is a liquid and not a glass. In fact,
a slightly different approximation due to Milchev \cite{68} renders the
entropy nonnegative at all temperatures, but deviates now a bit from
the simulation data in the other direction. Thus, these investigations
show that although Eq.~(\ref{eq12}) does indeed hold it does not imply
anything about the validity of the Kauzmann ``entropy catastrophe''.

\section{Can one map coarse-grained models onto atomistically realistic
ones?}

From the above comments it is clear that in simulations of simplified
coarse-grained models the range of times one can span is much larger than
the one for chemically realistic models that include atomistic detail
(microseconds rather than nanoseconds).  On the other hand, the simplified
models may elucidate general concepts but they fail to make quantitative
predictions on the properties of particular materials. Thus the question
arises whether one can somehow combine the advantages of both approaches.

An idea to do this is to make the coarse-graining process in a more
systematic way and to construct coarse-grained models that ``remember"
from which atomistic system they come from. For a polymer chain,
coarse-graining along the backbone of the chain may mean that if
we label the covalent bond consecutively $(1,2,3,4,5,6, \dots)$ the
bonds $1,2,3$ form the effective bond $I$, the bonds $4,5,6$ form the
effective bond $II$, etc \cite{69}. The potentials on the atomistic scale
(e.g. potentials controlling the lengths of covalent bonds, the angles
between them, the torsional angles, etc.) have then to be translated
into suitable effective potentials for the length $l$ of the effective
bonds and the angle $\Theta$ between them. The simplest choice would be
to assume potentials of the form

\begin{equation} \label{eq13}
U_{\rm eff} (l)=\frac{1}{2}u_0(l-l_0)^2 \, \, \, , V_{\rm eff}(\Theta)= \frac{1}{2}
\upsilon_0(\cos \Theta - \cos \Theta_0)^2 \quad.
\end{equation}

In the past potentials of this type have indeed be extracted from the
probability distributions $P(l) \propto \exp [-U_{\rm eff} (l)/k_BT] \, \,
\, , \,\, P(\Theta)  \propto \exp[-V_{\rm eff} (\Theta)/k_BT]$ observed
in the simulations of single chains (where long range interactions
need to be truncated, however) \cite{69,70}. Of course, the effective
parameters $u_0, l_0, \upsilon_0, \Theta_0$ are somewhat temperature
dependent, and in principle one should deduce them from simulations of
atomistically described melts containing many chains, rather than from
single-chain simulations \cite{70}. The practical implementation of how
one constructs best the effective potentials that mimic one particular
material is still an active topic of research \cite{44,70}.

\begin{figure}
\vspace*{3mm}

\centerline{
\psfig{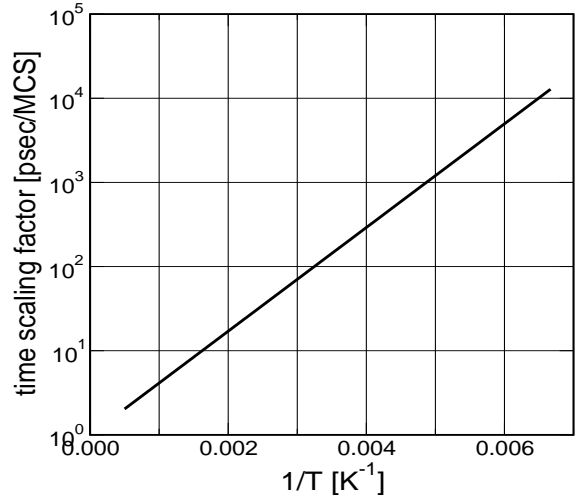}
}
\caption{\label{fig17}Temperature dependence of the time scaling factor
converting the time unit of the Monte Carlo simulation into femtoseconds,
for the case of polyethylene. The straight line shows that a simple
Arrhenius law is a good approximation. Adapted from Ref. \cite{44}.}

\end{figure}

A further important aspect is the question to what extent the
dynamics with such a coarse grained system reflects the dynamics
of a real chain. Here one needs to focus on the slowest local process,
which are hops of small groups of monomers to a new conformation, such
that a barrier of the torsional potential is crossed. Without such moves
involving barrier crossing no conformational changes can occur. In a
typical case, e. g. for polyethylene at $T=500$K, the time scale for
such hops is about two orders of magnitude larger than the vibration
times of bond lengths and bond angles. Only because of this separation
of time scales one can hope that a coarse-grained model can describe
the essential features of the slow dynamics in the polymer melt at all,
if the time units are properly rescaled. As shown by Tries {\it et al.}
\cite{44}, the knowledge of the torsional potentials allows, using a an
approach that resembles transition state theory, to construct a ``time
rescaling factor'', that gives the translation of the time unit of the
Monte Carlo simulations (attempted Monte Carlo steps per monomer) into
physical time units (Fig.~\ref{fig17}). One sees that for polyethylene
1 Monte Carlo step corresponds to 0.1 to 10ps, in the temperature region
of interest. At high temperatures, namely for $T=509$K, the accuracy of
the coarse-grained model of C$_{100}$H$_{202}$ was tested by running
a molecular dynamics simulation of a united atom model for about a
nanosecond (which is of the order of the Rouse relaxation time at this
temperature) for comparison \cite{44}. It is found that the agreement
between both approaches is almost quantitative. The advantage of the
Monte Carlo simulation of the coarse-grained model is, however, that one
can easily study a supercooled melt also at $T=250$K, a temperature which is
basically inaccessible to the molecular dynamics approach.

If one compares the results of the coarse-grained model to experimental
data, e. g. for the viscosity and its temperature dependence, the
agreement is encouragingly good but not perfect \cite{44}. One aspect
which is clearly missing in the coarse-grained model is the description
of attractive intermolecular forces. Thus, while this approach of mapping
atomistic models to coarse-grained ones clearly has a great potential,
there are still nontrivial problems that need to be solved.

\section{Concluding remarks}

In this brief review, the ``state of the art'' of computer simulations
of glassy systems was summarized. The main problem in this field is
the problem of bridging time scales - a supercooled fluid close to the
glass transitions exhibits a nontrivial dynamic behavior that extends
from very fast processes (in the picosecond time scale range) to very
slow processes (with relaxation times of the order of hours). Atomistic
molecular dynamics simulations of chemically realistic models (as
exemplified here for the case of molten SiO$_2$) can treat only a
very small part of this broad range of time scales, and also special
techniques such as the parallel tempering method can add only one or
two decades to this range but not more. (Note also that there are still
some unsolved technical problems with this method~\cite{demichele01}).
While such atomistic simulations are nevertheless useful, in particular
since they complement the time range directly accessible to experiment,
and give a very detailed insight into the interplay between structure
and dynamics in supercooled fluids, they clearly cannot answer questions
on the nature of relaxation processes for temperatures close to (the
experimental) $T_g$, and the possible existence for an underlying static
phase transition (from a metastable supercooled fluid to a metastable
ideal glass) at a temperature $T_K<T_g$. Also molecular dynamics studies
of coarse-grained models for melts of short, unentangled polymer chains
suffer from similar problems, although the effective cooling rates in
these models are about a factor of $10^3$ smaller than in the model
for silica, and one can access relaxation times that are almost in the
microsecond range. These models are very useful as a testbed for the mode
coupling description of the glass transition in fragile glassformers,
however. Furthermore they have also allowed to gain very useful insight
on the relaxation between the local motions responsible for the glass
transition (cage effect etc.) and the more mesoscopic Brownian motion
of the polymer chains (as described by the ``Rouse modes'', for instance).

A slightly more abstract model of the same systems, the bond fluctuation
model of glassforming polymer melts, corroborates these conclusions,
although due to its discrete nature it is somewhat less suitable to
describe the local structure of packing effective monomers in a polymer
melt or their motion on small scales (confined in a cage). However, this
model has the merit that it allows to compute the temperature dependence
of the configurational entropy $S(T)$ and thus to test the correctness
of theories like the one of Gibbs and Di Marzio. While it is found that
the entropy $S(T)$ decreases significantly if the polymer melt approaches
the glass transition, there is clear evidence that the theory of Gibbs
and Di Marzio is quantitatively very unreliable since it underestimates
$S(T)$ significantly at all temperatures, and the ``entropy catastrophe''
that it predicts is clearly an artifact of inaccurate approximations.

Finally, studies of an even more abstract model were discussed, the
$10$-state Potts glass with mean field infinite range interactions. This
model has the advantage that it is known exactly that it has a dynamical
(ergodic to nonergodic) transition at $T_D$ as well as a static transition
at a (slightly) lower temperature $T_0$, at which a glass order parameter
appears discontinuously and the entropy shows a kink.  The conceptional
disadvantage of this model, however, is that it has a built-in quenched
random disorder (via its random exchange couplings) at all temperatures,
unlike systems that undergo a structural glass transition, which have no
quenched disorder in the high temperature phase (the supercooled fluid
for $T>T_g$). Monte Carlo studies of this model, intended to serve as
a general testbed for systems with both a dynamical and a static glass
transition, show that unexpectedly large finite size effects occur,
which are poorly understood.  Thus even for this ``simple'' model much
more work is necessary.

While the anticipated progress in computer hardware and algorithmic
improvements will allow to extend the time ranges accessible in all
these simulations somewhat, there is not real hope that one can bridge
the desired $15$ (or more) decades in time in this way. More promising
in principle is the approach of providing an explicit mapping between
atomistic models (which cover the fast processes) and coarse-grained
models (which describe the somewhat slower processes, in the $10$ps to
$1\mu$s range), so that one effectively considers the same model system
but with different approaches on different time scales. Of course, this
idea is difficult to work out consistently in practice, and only modest
first steps towards its realization have been taken. Much more work in this
direction is certainly very desirable in the future.

\bigskip

\textbf{Acknowledgements}: We are particularly grateful to C. Bennemann,
C. Brangian, J. Horbach, T. St\"uhn, K.  Vollmayr, and M. Wolfgardt for
their valuable collaboration on parts of the research described here, and
acknowledge financial support from the Deutsche Forschungsgemeinschaft
(DFG/SFB 262), the Bundesministerium f\"ur Bildung und Forschung (BMBF
grant No 03N6015) and SCHOTT Glas. We thank the NIC J\"ulich and the
HLRS Stuttgart for generous allocations of computer time.

\vspace*{-20mm}

\vspace*{-25mm}

\end{document}